\documentclass[twocolumn,showpacs,preprintnumbers,amsmath,amssymb]{revtex4}
\usepackage{graphicx}
\usepackage{amssymb}
\usepackage[latin1]{inputenc}
\usepackage{dcolumn}
\usepackage{bm}

\begin{document}

\title{Temperature Induced Spin Density Wave   in Magnetic Doped
Topological Insulator Bi$_2$Se$_3$}
\author{Martha Lasia and  Luis Brey}
\affiliation{Instituto de Ciencia de Materiales de Madrid, (CSIC),
Cantoblanco, 28049 Madrid, Spain}

\date{\today}


\begin{abstract}
We study the magnetic properties of  Bi$_2$Se$_3$  doped with
isoelectronic magnetic impurities. We obtain that at zero
temperature the impurities order ferromagnetically, but when raising
the temperature the system  undergoes a first order phase transition
to a spin density wave phase  before the system reaches the
paramagnetic phase. The origin of this  phase is the non-trivial
dependence of the  spin susceptibility on the momentum. We analyze
the coupling of the non-uniform magnetic phase with the Dirac
electronic system that occurs at the surfaces of the topological
insulator.

\end{abstract}

\maketitle


\section{Introduction}
Topological insulators (TI) are a newly discovered type of systems
which are insulating in the bulk and characterized by the existence
of  a robust helical gapless Dirac two dimensional electron system
at their surface\cite{hasan_2010,Qi_2010,Qi_2011}.

TI's are typically band insulators for which strong spin orbit
coupling produces an inversion of the bulk band gap. Therefore, in
TI's the energy gap is related with the spin orbit coupling and that
limits its  magnitude. The most studied and more promising
topological insulator is Bi$_2$Se$_3$, which is a three dimensional
TI with a relatively large  bulk energy gap $\sim 0.3$eV and with
the Dirac point of the surface states located outside the bulk
bands\cite{Xia_2009,Zhang_2009}. Angle resolved
spectroscopy\cite{Hsieh_2009,Xia_2009} and scanning tunneling
microscopy\cite{Hanaguri_2010}experiments have shown the Dirac
nature of the surface states of Bi$_2$Se$_3$.

The spin and wavevector of the surface states of a TI are  strongly
coupled, and  the occurrence of a half-quantized Hall effect when a
energy gap opens at the surface has been
predicted\cite{Qi_2008,Essin_2009}. Due to the protected character
of the Dirac states, a gap at the surface should be opened with a
perturbation that breaks the time reversal symmetry. This can be
done by doping the system with magnetic impurities. At the surface
of the TI, because of the large spin-orbit coupling, the interaction
between the  Dirac-like surface states and the impurities induces a
large single ion magnetic anisotropy and polarizes the spin of the
impurities perpendicularly to the surface. This spin-orbit coupling
translates in the opening of an energy gap at the Dirac point of the
surface
states\cite{Liu_2009,Biswas_2010,Yu_2010,Jiang_2011,Yokoyama_2011,Zhu_2011,Abanin_2011,Menshov_2011,nunez_2012,Tetsuro_2012}.

From the experimental side, angle resolved photoemission
spectroscopy (ARPES) studies on the surface of Fe-doped Bi$_2$Se$_3$
single crystals have confirmed the opening of an energy gap at the
Dirac point\cite{Chen_2010} and the creation of odd multiples of
Dirac fermions\cite{Wray_2010}. Also, recently, experiments in thin
films of Cr-doped Bi$_x$Sb$_{2-x}$Te$_{3}$,  has shown a large
anomalous Hall conductance in a magnetically doped topological
insulator\cite{Chang_2011}.

However, recent experiments\cite{Honolka_2011} found that the spins
of Fe ions deposited on Bi$_2$Se$_3$ orient  inplane. Also ARPES
experiments\cite{Scholz_2011,Valla_2012} found Dirac crossing even
in the presence of magnetic impurities in contradiction with earlier
experiments and existing theory. On the other hand, recently it has
been reported the suppression of the Dirac point spectral weight,
both in magnetically doped and undoped TI, suggesting that the
observed gap at the  Dirac point can not be taken as the sole
evidence of a magnetic gap\cite{Xu_2012}. In addition, density
functional theory based calculations\cite{Schimdt_2011} find that Co
adatoms lying in the Bi$_2$Se$_3$ surface exhibit an energetically
stable magnetic moment perpendicular to the surface, whereas  for Co
atoms located on the interlayer Van der Waals spacing the momentum
is in the plane parallel to the surface. All these results indicate
the complexity of the interpretation of the ARPES experiments and
the possible importance of other effects not included in the Dirac
hamiltonian, as crystalline anisotropy or surface reconstruction,
might play an important role on the orientation of the magnetic
impurities. In this work we use an effective hamiltonian for
describing Bi$_2$Se$_3$, which although it does not include
microscopic details of the material describes appropriately the
basic properties of the Bi$_2$Se$_3$ related with its band structure
topology.

In this work we study the phase diagram of   magnetically doped
Bi$_2$Se$_3$. Bi$_2$Se$_3$ is a layered  material formed by five
atom layers arranged along the $z$-direction. We find that at low
temperatures the magnetic impurities order ferromagnetically along
the $z$-direction. By raising the temperature, the TI undergoes two
transitions; A first order transition from the ferromagnetic to the
spin density wave phase, and at higher temperatures  a  second order
transition from the spin density wave phase to the paramagnetic
phase. The spin density wave phase has both,the polarization and the
wavevector, parallel to the $z$-direction. We have also studied the
effect of the surface states by calculating the magnetization as
function of temperature of a slab of Bi$_2$Se$_3$ topological
insulator. Here we find that the surface magnetization survives to
higher temperatures than the bulk spin density wave phase.

The paper is organized as follow, in Section II we define the
hamiltonian we use for describing the electrical properties of
Bi$_2$Se$_3$. In Section III we calculate the wavevector dependent
paramagnetic spin susceptibility of Bi$_2$Se$_3$ and discuss the
interaction between magnetic impurities through the paramagnetic
susceptibility. In Section IV we formulate  a Landau theory for
describing the magnetic order of magnetically doped Bi$_2$Se$_3$,
and discover the existence of a ferromagnetic to spin density wave
phase transition at finite temperature. In Section V we study the
polarization profiles of a magnetically doped Bi$_2$Se$_3$ slab and
analyze the effect that the Dirac-like surface states have on the
magnetic phases. We finish in Section V with some conclusions and
remarks.

\section{Hamiltonian}
The low energy and long wavelength electronic properties of
Bi$_2$Se$_3$ topological insulators are described by the four bands
${\bf k }\cdot {\bf p }$ Hamiltonian\cite{Zhang_2009},
\begin{eqnarray}
& & H     \! \!  =    E({\bf k})    \label{H3D} \\
  & +&     \mathcal{M} ( {\bf k} ) \tau _z \otimes  I  \! + \!   A_1 k_z  \tau _x \otimes \sigma _z  \! + \!
     A_2 (k _x \tau _x + k_y \tau _y) \otimes \sigma _x  \nonumber
     \label{H0}
\end{eqnarray}
where $\sigma _{\nu}$ and $ \tau _{\nu}$ and Pauli matrices, $I$ the
unity matrix, $\mathcal{M} ( {\bf k} )$=$M_0 - B_2 (k_x ^2 + k_y
^2)-B_1k_z ^2$, $k_{\pm}$=$k_x \pm i k_y$ and  $E({\bf k})$=$ C + D
_1 k_z ^2 + D_2 (k_x ^2+k_y ^2)$. The Hamiltonian is written in the
basis $|1>$=$|p1 _z ^+, \uparrow>$, $|2>$=$-i |p2 _z ^- ,
\uparrow>$,$|3>$=$|p1 _z ^+, \downarrow>$, $|4>$= $i|p2 _z ^-,
\downarrow>$, which are the hybridized states of the Se-$p$ orbital
and the Bi-$p$ orbital with even $(+)$ and odd $(-)$ parities and
spin up ($\uparrow$) and down ($\downarrow$).
The Hamiltonian parameters for Bi$_2$Se$_3$
are\cite{Liu_2010}
, $M_0$=0.28$eV$, $A_1$=0.22$eV nm $, $A_2$=0.41$eVnm$,
$B_1$=0.10$eV nm ^2$, $B_2$=0.566$eVnm^2$, $C$=-0.0068$eV$,
$D_1$=0.013$eV nm ^2$ and $D_2$=0.196$eVnm^2$.
In this basis the spin operators get the form\cite{Silvestrov_2011},
$ S_z = I \otimes \sigma _z , \, \, \, S_x = \tau _z  \otimes \sigma
_x \,\mathrm{ and}  \, \, S_y = \tau _z \otimes \sigma _y $.

\section{Bulk Spin Susceptibility}
The paramagnetic susceptibility obtained from the Hamiltonian
Eq.\ref{H3D} has the form
\begin{equation}
\chi _{\mu \mu} ({\bf q}) \! = \!  \frac 2 {\Omega}   \!  \sum _ {
\stackrel { {\scriptstyle n'  \,  \mathrm{occ.} }}
          { {\scriptstyle  n   \, \mathrm{empty} }}}
\sum _{ {\bf k }}  \frac {  | \! < \!  n', {\bf {k+q}}| S _{\mu} |n,
{\bf k} \! > \! |^2  } { \varepsilon _{n' ,{\bf k}+{\bf q}} -
\varepsilon _{n ,{\bf k}}} \! . \label{chimumu}
\end{equation}
Here $|n, {\bf  k}>$ and $\varepsilon _{n ,{\bf k}}$ are the
eigenfunctions and eigenvalues of Hamiltonian Eq.\ref{H3D} and
$\Omega$ is the sample volume.
In the case of an insulator, this spin susceptibility is caused by
the coupling of the valence and conduction band induced by the spin
operator\cite{Vleck_1932}. The susceptibility is a smooth function
of the wavevector and because the system is an insulator there are
no anomalies associated with Fermi surfaces. The symmetry of the
original Hamiltonian dictates that the non-diagonal elements of the
susceptibility tensor are zero and $\chi_{xx}= \chi_{yy} \ne
\chi_{zz}$.

\begin{figure}
 \includegraphics[clip,width=8.cm]{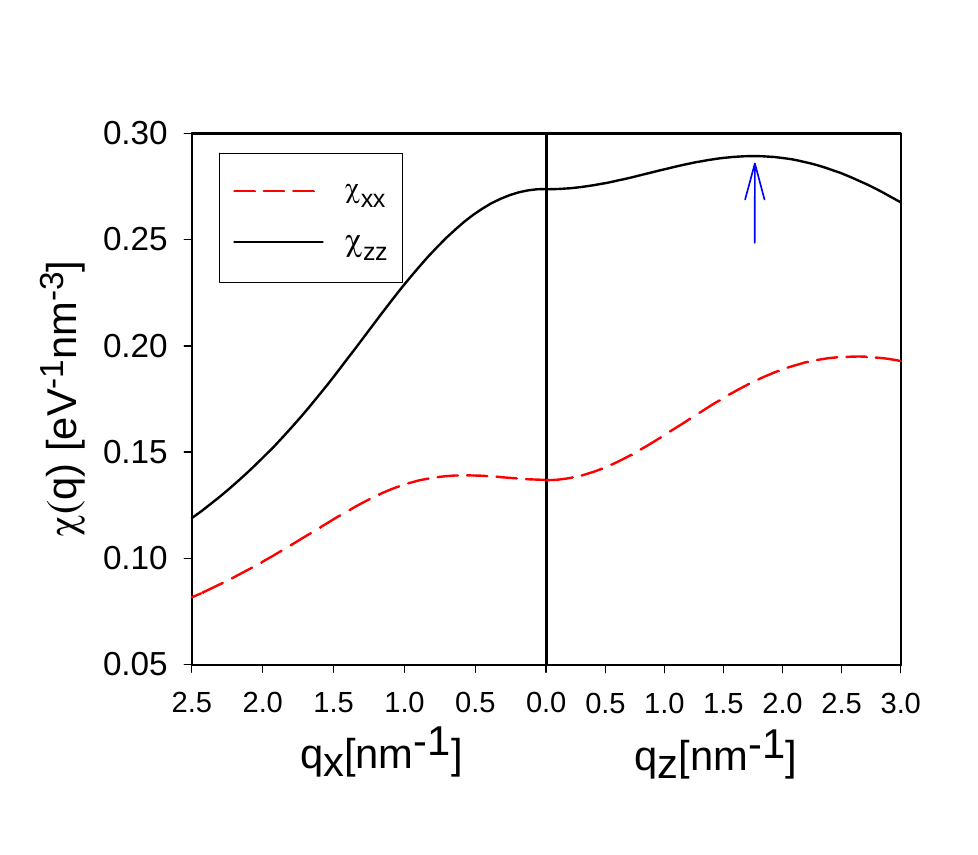}
 \caption{ (Color online) Spin susceptibility as function of the wave vector along the $z$ and the $x$-directions.
 The arrow indicates the position of the maximum.}
  \label{chi_qxqz}
\end{figure}

In Fig.\ref{chi_qxqz}  we plot the $\chi _{xx}$ and $\chi_ {zz}$ as
a function of $q_z$ and $q_x$. The direct coupling, $A_2 k_{\pm}$,
between atomic orbitals with opposite parities and opposite
$z$-component of the spin, makes that for $k_{\pm} \ne 0$, occupied
and empty states are coupled through $S_z$. Whereas those states are
only connected through   $ S_x $   when $k_z \ne 0$. This makes
$\chi _{zz} ({\bf q}) > \chi_{xx} ({\bf q})$.


The more important contribution to $\chi _{zz} (q_z )$ comes from
regions in the reciprocal space where the matrix elements $ < \! n',
{\bf {k}}+q_z| S _{z} |n, {\bf k} \! >|$, with $n$ occupied and $n'$
empty, reaches the maximum value. This happens   when $\mathcal{M} (
{\bf k} )$=$0$ or    $\mathcal{M} ( {\bf k}+ q_z)$=$0$. For a given
$k_z$ these conditions define two  circular crowns
of radius $\sqrt{\frac {M_0 - B_1 k_z ^2}{B_2}}$ and $\sqrt{\frac
{M_0 - B_1 (k_z + q_z)^2}{B_2}}$ and thickness $A_2 /(2 B_2)$.
Therefore, the area  of the reciprocal space that contributes
appreciably to  $\chi_{zz} (q_z)$ increases with $q_z$. For larger
values of $q_z$ one of the circular crowns collapses to zero and the
contributions to the integral decrease. This behavior explains
qualitatively the maximum that $\chi _{zz}$ presents at a wavevector
$G \sim  \sqrt{M_0 / B_1}$.

\begin{figure}
 \includegraphics[clip,width=8.cm]{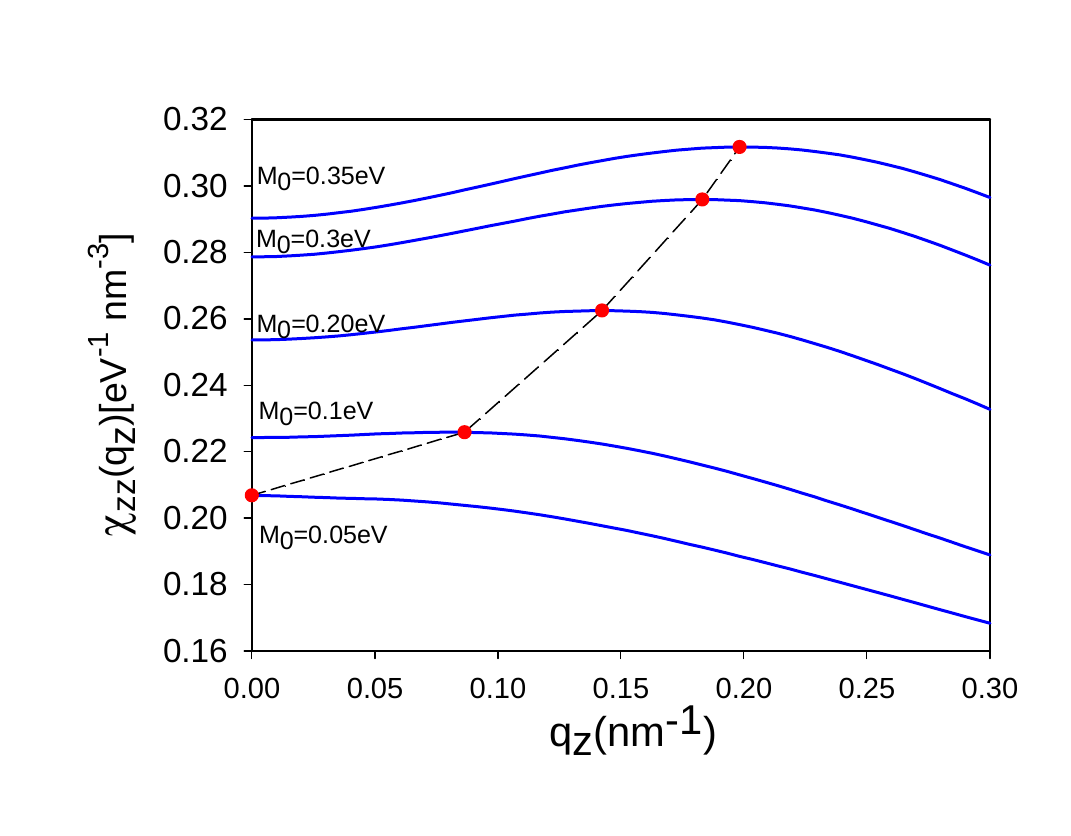}
  \caption{(Color online)Spin magnetic susceptibility, $\chi_{zz}$,  for different values of $M_0$, as function of the momentum in the $z$-direction. All other parameters of the band structure corresponds to those of Bi$_2$Se$_3$.
 As the value of the mass $M_0$ decreases the position of the maximum of the susceptibility
 moves towards small values of $q_z$. At small values of $M_0$ the maximum occurs at $q_z=0$.
 In the normal insulator phase, $M_0 <0$,
 the maximum of   $\chi_{zz}$ occurs at $q_z=0$ for all values of the mass parameter.
 The dots indicate the position of the maximum.}
  \label{chizz_M0}
\end{figure}

The existence of a maximum in  $\chi _{zz} ({\bf q_z})$ at finite
$q_z$ is robust against small changes in the parameters of the four
bands Hamiltonian. In Fig.\ref{chizz_M0} we plot $\chi _{zz} ({\bf
q_z})$ for different values of the TI gap. The position of the
maximum decreases continuously towards $q=0$ when $M_0$ decreases
and only disappears  for small values of $M_0$. In the normal
insulator phase, $M_0 <0$, the maximum always occurs at $q=0$.

\subsection{Coupling between diluted magnetic impurities}
Consider now a TI doped with magnetic impurities of spin $S$. We
assume that the number of electrons in the system does not change in
the presence of the magnetic impurities. That can be achieved by
doping with isoelectronic magnetic dopants or by adding compensating
non-magnetic dopants\cite{Hor_2009}.  In this work we consider the
dilute limit, i.e. concentration of impurities smaller than 5$\%$,
for which the direct interaction between the spin of the magnetic
impurities can be neglected.

However, the electrons spins  have a strong exchange coupling,
$\frac S 2 \tilde{J}(|{\bf r}|)$, with the magnetic impurities spins
which, in turn, are equally affected by the exchange field of the
electrons. In this form  the magnetic impurities in the system
interact mediated by electronic states.  We treat this interaction
in second order perturbation theory\cite{Dietl_2000,Brey_2003}, that
has proved to be a reliable approximation in diluted magnetic
semiconductors\cite{Brey_2003,Calderon_2002}. In this approach the
effective exchange parameter between two magnetic impurities
separated by a vector ${\bf R}$ and spins pointing in the
$\nu$-direction is,
\begin{equation}
J _{\nu}({\bf R})=-\frac {S^2} 4{J_{eff}^2 }{\Omega} \sum _{\bf q}
\chi _{\nu \nu} ({\bf q}) e ^{i {\bf q}{\bf R}}
\end{equation}
where $J_{eff}= \int \tilde {J} (|{\bf r}|) d {\bf r}$ is the
effective exchange coupling between the magnetic impurity and the
electron spin.

Because $\chi _{zz} > \chi_{xx}$ in all range of wavevectors, the
system has an easy axis of magnetization along the $z$-direction and
therefore isoelectronic magnetic impurities in Bi$_2$Se$_3$ will
tend to polarize in the $z$-direction. The maximum that the spin
susceptibility presents at finite wavevector in the $z$-direction,
will determine the existence of non-uniform polarization in
magnetically doped TI. We treat the  magnetically ordered state  in
the virtual crystal
approximation\cite{Dietl_2000,Abolfath_2001,Brey_2003,FR_2001}, and
we consider that the system is invariant in the $(x,y)$-plane, and
the polarization only depends on the $z$-direction.
In the next section we obtain the magnetic polarization as function
of temperature and $z$-coordinate by using a Landau free energy
functional.

\section{Landau free energy functional}


We assume that the system is invariant in the $(x,y)$-plane, and
allows the polarization to oscillate with period $2\pi/G$ along the
$z$-direction. In consequence we define  the normalized magnetic
polarization, $-1 \le m(z,T) \le 1$, as,
\begin{equation}
m(z,T)= m_0(T)+m_G(T) \cos (G z) \, \, \, , \label{mzT}
\end{equation}
where  $m_0$ and $m_G$ are the order parameters of the uniform
ferromagnetic (FM) phase and the  spin density wave phase (SDW)
respectively.

The internal energy per unit volume corresponding to this
magnetization is,
\begin{equation}
E= - \frac J 2 m_0 ^2   \chi _{zz} (0) -  \frac J 4 m_G ^2   \chi
_{zz} (G) \, \,.
\end{equation}
where $J= \frac {S^2} 4 J ^2 _{eff}  c$, being $c$ the density of
magnetic impurities. In our case, the value of $\chi _{zz} (G) $ is
less than 10$\%$ larger than $\chi _{zz} (0) $ and the zero
temperature ground state is  a uniform FM phase, $m(z,T=0)=1$.
However the maximum of the spin susceptibility at $G$ will modify
the spin density at larger temperatures.

Knowing that for small values of the polarization, the entropy of a
classical spin at a given $T$ is, see Appendix A,
\begin{equation}
-TS=-k_B T \ln(2) + \frac 3 2 k_B T m ^2+ \frac 9 {20} k_B T m^4 \,
\,
\end{equation}
we get that in the mean field approximation and for small values of
magnetic polarization, the Landau free energy per unit volume takes
the form,
\begin{eqnarray}
\mathcal {F} = - \frac J 2 m_0 ^2    \chi_{zz} (0)- \frac J 4 m_G ^2    \chi_{zz} (G) \nonumber \\
- \frac 1 {\beta} \frac 1 L \int dz \left \{ \ln 2 - \frac 3 2 m  ^2
(z,T)  - \frac 9 {20} m ^4 (z,T) ... \right \} \, \, ,
\end{eqnarray}
where $\beta = 1/k_B T$ and $L$ is the sample dimension in the
$z$-direction. Using the expression Eq.\ref{mzT}, and in the limit
$\L \rightarrow \infty$, we get
\begin{eqnarray}
\mathcal {F}   &  = &
\frac 3 2 m_0 ^2 k_B (T  -  T_0)  +   \frac 3 4 m_G ^2 k_B (T  -  T_G) \nonumber \\
 & + &  k_B T \frac {27}{20} m ^2 _0  m^2 _G  + k_B
T \frac {9}{20} m_0 ^4  +  k_B T \frac {27}{160} m_G ^4
\label{functional}
\end{eqnarray}
where $T_0$ and $T_G$ are the critical temperatures of the pure FM
and SDW phases respectively,
\begin{equation}
k_B T _0  \!  =  \! \frac {J}{3} \chi_{zz}(0)  \,  \, \, \mathrm{and
}\, \, \,  k_B T _G  \! =  \!  \frac {J }{3} \chi_{zz}(G) \, \, \, .
\end{equation}
The phase diagram of a system described by a  free energy as that of
Eq.\ref{functional} depends on the relative magnitudes of the
fourth-order potentials\cite{chaikin_book}. In our case the product
of the pre-factors of $m_0 ^4$ and $m_G ^4$ is smaller than the
square of the    $m_0^2m_G^2$  pre-factor and there is no phase
coexistence in the phase diagram. By increasing the temperature,
there is a first order transition from the FM phase to the SDW phase
at
\begin{equation}
T ^* = \frac {\sqrt{3} T_0 - \sqrt{2} T_ G}{\sqrt{3} -\sqrt{2}} \, \, .
\end{equation}
This is the main result of this work: by heating, a magnetically
doped TI undergoes two phase transitions, a FM to SDW first order
transition at $T ^*$ and  a SDW to paramagnetic second order
transition at $T_G$. Although at $T$=0, the FM phase has lower
energy than the SDW phase, the FM to SDW transition at finite $T$
occurs because the entropy of the SDW increases  faster with $T$
than the  entropy of the FM phase.

In the next section we analyze how the surface states existing in
topological insulators couple to the bulk magnetic polarization.

\section{Spin polarization of magnetically doped TI slabs}
At the surface of a  TI there exists a two dimensional Dirac
electron gas. Because the chirality of the electron gas, an exchange
field perpendicular to the surface opens a gap in the spectra. Then,
in order to minimize the energy, a magnetic impurity will polarize
perpendicularly to the
surface\cite{Liu_2009,Biswas_2010,Yu_2010,nunez_2012}. In the
diluted limit, surface states  mediate an RKKY interaction among the
impurities  which is always ferromagnetic, whenever the chemical
potential resides near the Dirac
point\cite{Liu_2009,Yu_2010,nunez_2012,Brey_2007}. Therefore
magnetic impurities at the surface of a TI will order
ferromagnetically perpendicular to the surface.

We are going to study numerically the spin polarization as function
of temperature and position of a magnetically doped TI slab. The
objective here is  first to confirm the results obtained with the
Landau functional where we consider  a unique Fourier component of
$\chi_{zz} (q_z)$ and second to analyze the coupling between the
surface and the bulk magnetization.

We analyze  a TI slab of thickness $L$ and perpendicular to the
$z$-direction. We expect the  electron affinity of Bi$_2$Se$_3$ to
be much larger than its band gap. Therefore, at the surface of the
TI we will neglect the penetration of the electron wavefunction into
the vacuum. The eigenvalues,  $\varepsilon _{n,{\bf k}}$, and
wavefunctions, $\Psi _{n, {\bf k}} (z)$, are obtained by solving
Eq.\ref{H3D} with $k_z = -i\partial _z$ and forcing the wavefunction
to vanish at $z=0$ and $z=L$. This is satisfied expanding $\Psi _{n,
{\bf k}} (z)$ in harmonics,
\begin{equation}
\Psi _{n, {\bf k}} (z)= \frac {e ^{i {\bf k } {\bf r}}} { \sqrt{A} }
\sqrt { \frac 2 L}\sum _{l =1} ^{N_{\mathrm{max}}}  \sum _{ j=1,4 }a
_{n,j} ^l ({\bf k}) \sin  {( \frac l L  \pi z)} \, \, ,
 \label{2Dwf}
\end{equation}
here $A$ is the sample area and we choose  $N_{max}$ large enough so
that the results does not depend on it.

\begin{figure}
 \includegraphics[clip,width=8.cm]{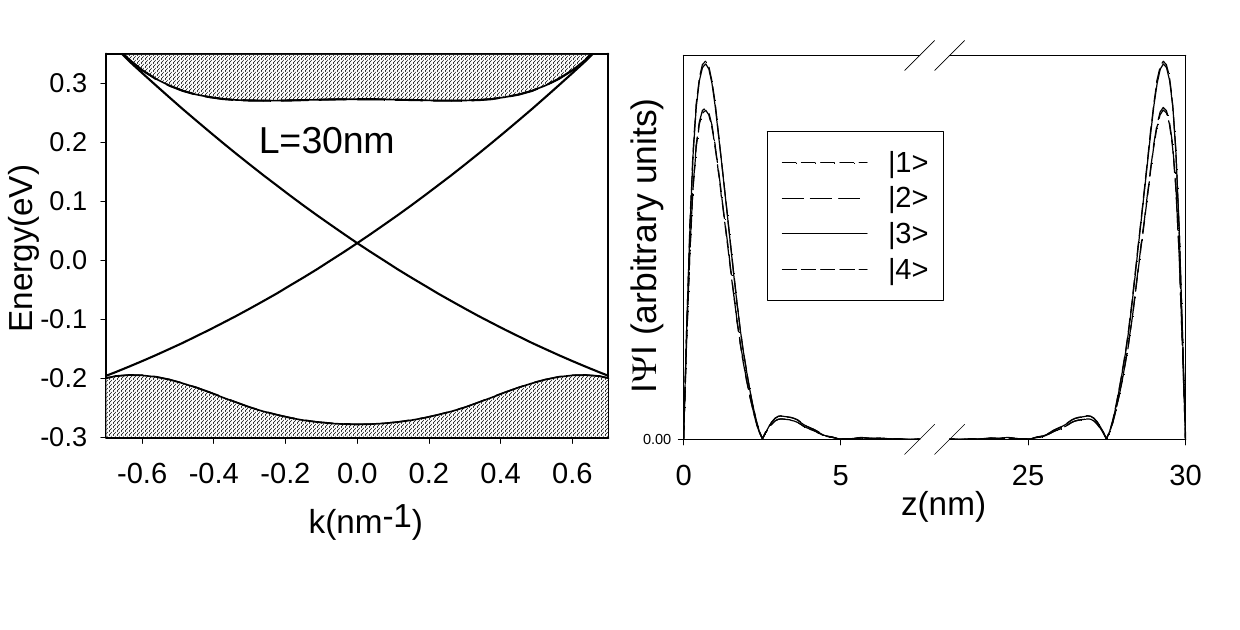}
 \caption{ (Color online)In the left panel we plot the band structure of
 a topological insulator film
 30$nm$ thick. For this thickness there is no coupling between localized states
 on opposite surfaces and surface states are degenerated.
  Dashed areas denote the bulk  band structure.  In the right panel we plot the
  absolute value of the four components of the wavefunction of a
surface state with momentum close to zero. }
  \label{surface_bands_wf_mandar1}
\end{figure}

For $L>10$nm, the surfaces of the slab are decoupled and the band
structure is independent of $L$. In the bulk energy gap region,
appear some surface states which are the benchmark of the TI. In
Fig.\ref{surface_bands_wf_mandar1} we plot the   band dispersion and
the shape of the  wavefunction of these states. The results we
obtain agree completely with previous
results\cite{Qi_2011,Silvestrov_2011}.

\begin{figure}
 \includegraphics[clip,width=8.cm]{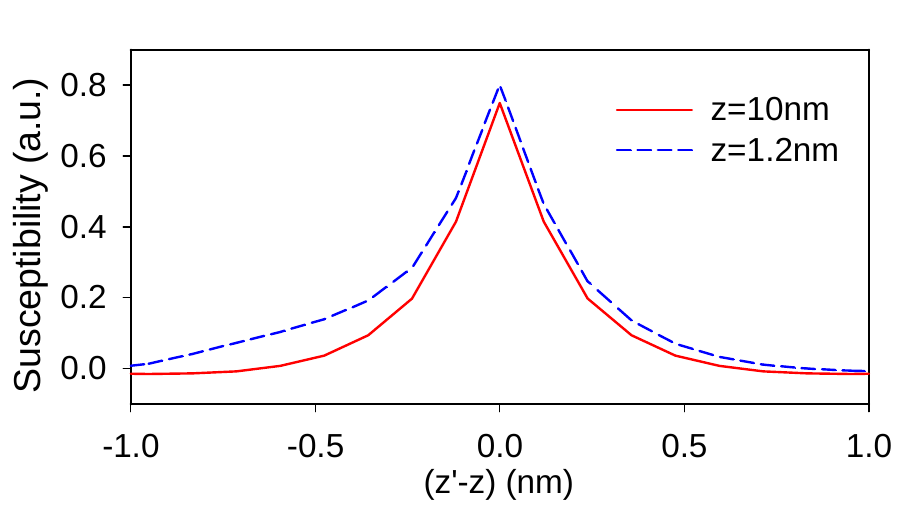}
 \caption{ (Color online)$\tilde{\chi} _{zz} (z,z')$
 evaluated at  the maximum of the surface
wavefunction, $z$=1.2nm,
 and at
 the center of a 20nm thick slab, as function of $z'$. The first case corresponds to a region near the surface, where the two dimensional
 Dirac electron system contributes to the response functions.
 In the latter case the response function is   not affected by the surface and it is the bulk response function. In both cases the functions are very peaked
at $z$=$z'$.  The negative values of the coupling in the bulk
response function is a consequence of the maximum that the response
function present at $q_z$=$G$ in the reciprocal space.  Near the
surfaces, and because of their metallic character, the magnetic
coupling is stronger. This is reflected in the asymmetry of the
dashed line, the interaction between planes is larger as closer the
planes are to the surface.}
  \label{chizz}
\end{figure}

In the slab geometry the momentum in the $z$-direction is not a good
quantum number and the paramagnetic susceptibility depends on two
position indices $z$ and $z'$. Therefore,  in the virtual crystal
approximation and in second order perturbation theory,  the internal
energy of the magnetically doped TI slab is,
\begin{eqnarray}
E =  \frac J {2 L} \!    \int _{0} ^{L} \! \! \!    \int _{0}
^{L} \! \!dz dz'    \tilde{\chi} ( z,z') m(z) m(z') ,\,  \, \,   \mathrm{with} \\
\tilde{\chi}  (z,z') =  \frac 1 A \sum _{n,n', {\bf k } }
 \frac { n_F ( \varepsilon _{n ,{\bf k}} )- n_F( \varepsilon _{n' ,{\bf {k}}})  }
 { \varepsilon _{n' ,{\bf {k}}} - \varepsilon _{n ,{\bf k}}}  \times \nonumber
 \\
 \Psi _{n, {\bf k}}  ^* (z)   S_z  \Psi _{n', {\bf k}} (z)  \times
 \Psi _{n ', {\bf k}} ^* (z')    S_z  \Psi _{n, {\bf k}}
 (z')  \, \, ,
 \label{chi_2d}
\end{eqnarray}
where $n_F(\varepsilon)$ is the Fermi distribution function.
$\tilde{\chi}  (z,z')$ indicates the coupling between uniform
polarized $(x,y)$-planes,  located at positions $z$ and $z'$. The
interaction between magnetic impurities is mediated by electrons in
the system, and because the bulk  system is an insulator, the
interaction is very short ranged in the $z$-direction, see
Fig.\ref{chizz}.

\begin{figure}
 \includegraphics[clip,width=8.cm]{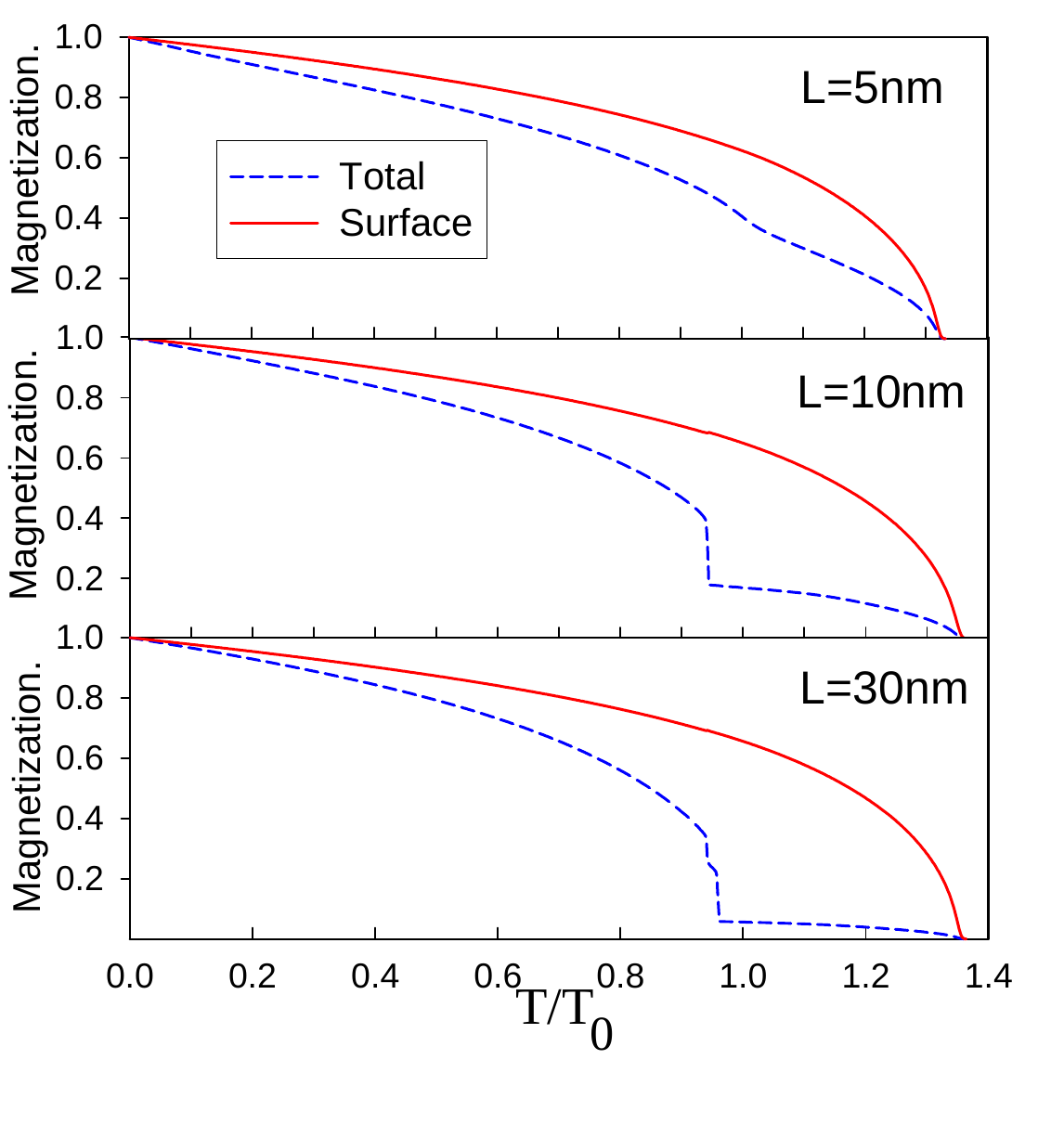}
 \caption{(Color online)Magnetization as a function of temperature for
 TI slabs of thickness (a) $L=5$nm, (b) $L=10$nm  and (c) $L=30$nm.  $T_0$ is the bulk FM critical temperature
of
 the topological insulator. The small "step" in the middle of the
 first order transition that occurs at $L=30$nm is consequence of
 an interference effect between the surface magnetization and the
 bulk SDW phase.
     }
  \label{Mag_T_L}
\end{figure}

\begin{figure}
\includegraphics[clip,width=8.cm]{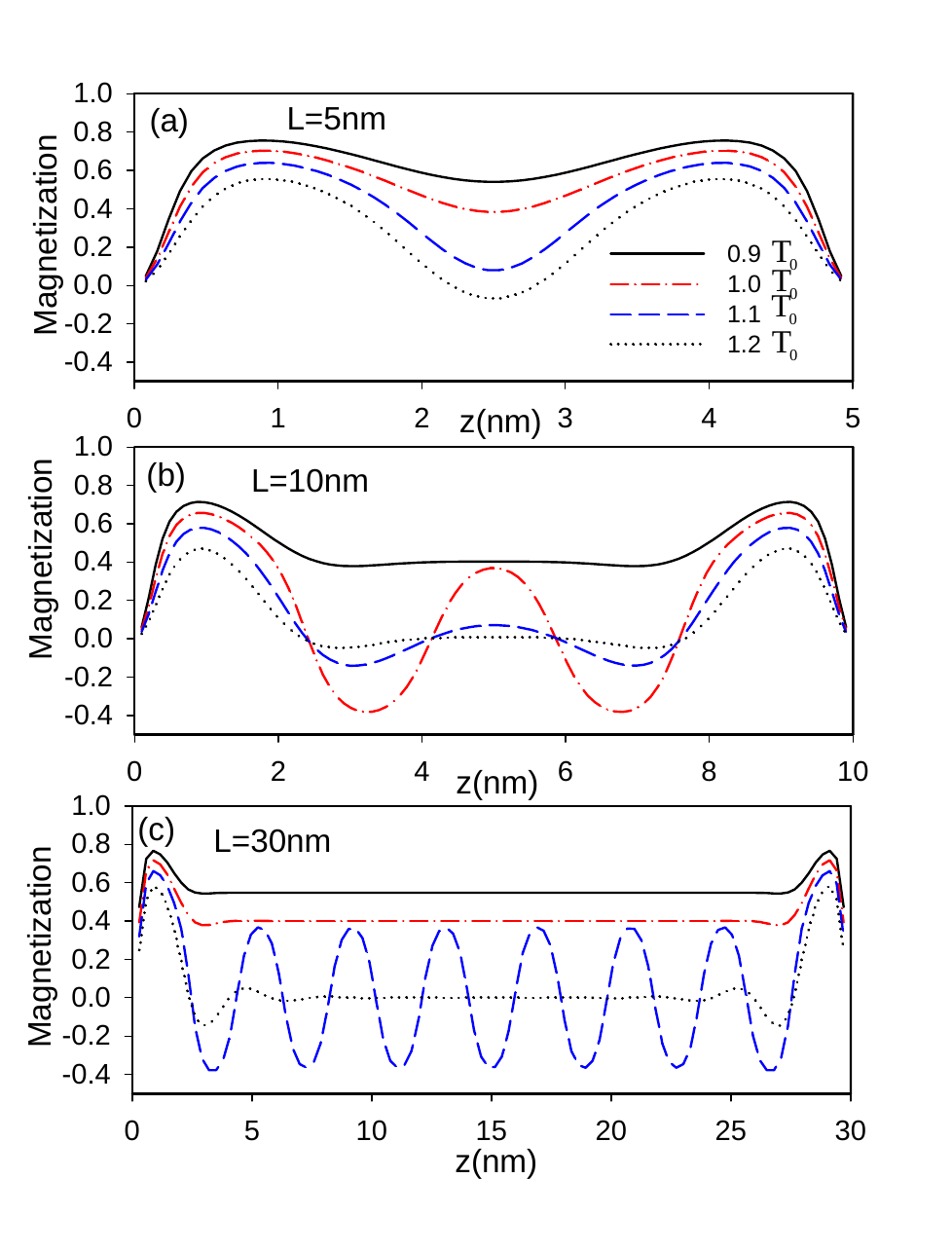}
 \caption{ (Color online)Magnetic polarizations as  a function of the position across the topological
 insulator slab $z$, for different layer thickness and temperatures.  }
  \label{MagTz_zz}
\end{figure}

We compute the temperature dependence of the magnetization profile
in the mean field approximation. At a given
position $z$, the magnetization $ m
(z,T)$ feels a (in energy units) magnetic field,
\begin{equation}
B(z) = J \int _0 ^L dz' \tilde{\chi} (z,z') m (z') \, \, ,
\label{beff}
\end{equation}
and  the magnetization of an isolated impurity in the presence of
the molecular field is,
\begin{equation}
m(z,T)= \coth \left (\frac {B(z)}{k_B T} \right )  -
\frac {k_B T}   {B }\, \,  . \label{mT}
\end{equation}
Solving self-consistently Eq.\ref{beff} and Eq.\ref{mT}, we obtain
the magnetization profiles as a function of T.

Because the metallic surface states intermediate a RKKY
coupling\cite{Brey_2007,Liu_2009} at the surface, the response
function, $ \tilde \chi(z,z')$ is larger near the surface than in
the bulk, see Fig.\ref{chizz}. Therefore, as function of $T$, the
absolute value of the magnetization decreases  faster in the bulk
region than in the surface\cite{Rosenberg_2012}. However, it is
important to note that the surface and the bulk  are part of a
unique system and therefore there is only a {\it unique critical
temperature}, corresponding to the transition of the paramagnetic
phase.

In Fig.\ref{Mag_T_L} we show the magnetization as function of
temperature for  TI slabs of thickness $L=5$nm, $L=10$nm and
$L=30$nm. We plot the average value of $m(z,T)$,  and the value of
the magnetization on top of the surface states. In Figure
\ref{MagTz_zz}, we plot the magnetization profiles for different
temperatures and $L=5$nm, $L=10$nm and $L=30$nm.

For $L=10$nm and $L=30$nm the surfaces are practically decoupled and
the central part of the slab behaves as bulk. There is a strong jump
in the magnetization at $T^*$, that indicates the first order FM to
SDW transition. In the SDW phase the oscillating magnetization does
not contribute to the total magnetization and the magnetization for
$T>T^*$ is due to surface states. In Fig.\ref{MagTz_zz}(b)-(c), it
is apparent at the center of the slab, the abrupt transition from an
uniform magnetization phase to a SDW phase. For smaller thickness of
the slab, Fig.\ref{MagTz_zz}(a), the surface states are coupled and
there is no well defined bulk region, that reflects in the absence
of FM to SDW transition.

The magnetization at the surface is practically not affected by the
FM to SDW transition, and decays with $T$ continuously to zero. The
ferromagnetism at the surface is more robust than in the central
part. For temperatures where $m_0$ and $m_G$ are near zero, the
surface of the system can be more than 30$\%$ polarized. These
results indicate the possibility that the magnetization at the
surfaces of TI's could be finite at temperatures larger than the
bulk critical  temperatures $T_G$ and $T^*$\cite{Rosenberg_2012}.
Because of the metallic character of the TI surface states, there is
a range of temperatures, for which the Dirac-like electron system at
the surface of the TI is gapped, although the bulk part of the
system is practically unpolarized.

A similar SDW phase has been also obtained numerically by Rosenberg
and Franz in a slab geometry of Bi$_2$Se$_3$\cite{Rosenberg_2012}.
However these authors interpret the oscillation of the polarization
as spatial fluctuations of the bulk magnetization coupled with the
surface magnetization. From our calculation we attribute the
oscillations in the magnetization reported in
ref.\cite{Rosenberg_2012} as a signature of the bulk SDW phase.

\section{Final Remarks and Conclusions}

In this work we study the phase diagram of   magnetically doped
Bi$_2$Se$_3$. At low temperatures the magnetic impurities order
ferromagnetically along the $z$-direction. By raising the
temperature, the TI undergoes two transitions; A first order
transition from the ferromagnetic to the spin density wave phase,
and at higher temperatures  a  second order transition from the spin
density wave phase to the paramagnetic phase. This results could
explain recent experimental results\cite{Salman_2012} that suggest
the existence, as function of the temperature, of two different
magnetic phases in Fe doped Bi$_2$Se$_3$.

We have also studied the effect of the surface states by calculating
the magnetization as function of temperature of a slab of
Bi$_2$Se$_3$ topological insulator. Here we find that the surface
magnetization survives to higher temperatures than the bulk spin
density wave phase. The existence of a range of temperatures for
which the bulk magnetization practically vanishes whereas a finite
magnetization exits at the surface, could explain some experimental
results that observe a gap at the  surface of Bi$_2$Se$_3$ but not
bulk magnetism\cite{Wray_2010,Chen_2010}.

\begin{figure}
 \includegraphics[clip,width=8.5cm]{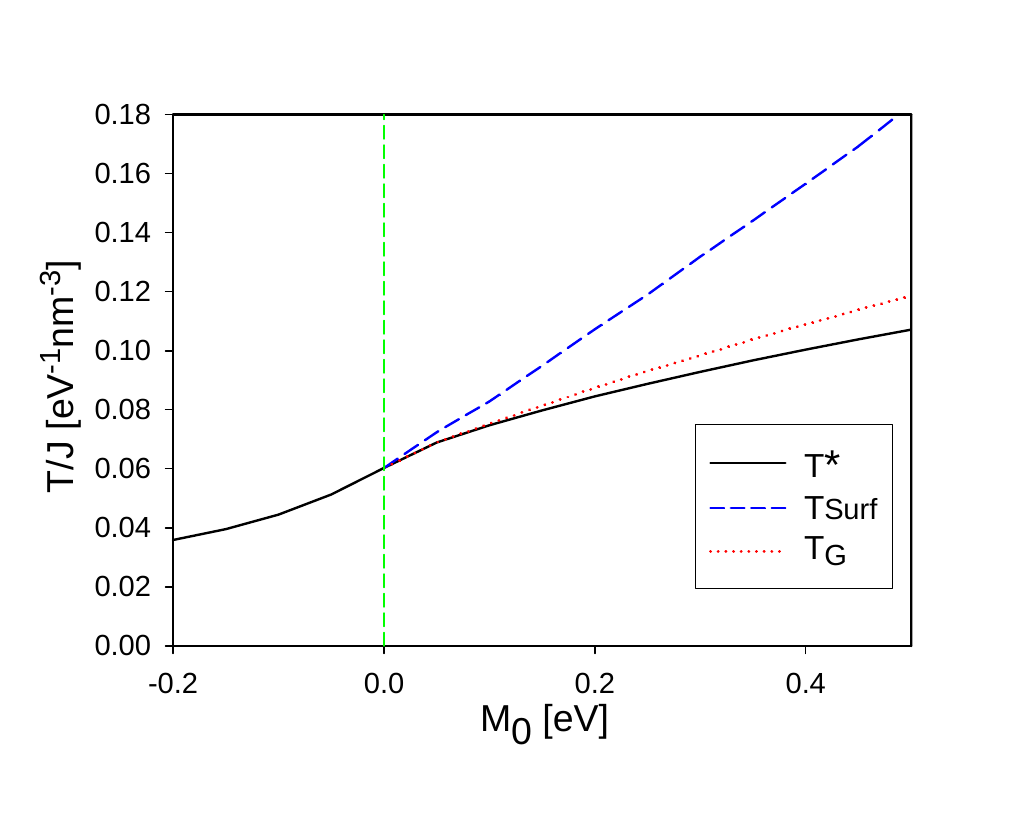}
 \caption{(Color online) Phase diagram of a thick magnetic doped TI,
 as function of the mass parameter $M_0$.
     }
  \label{Curie_Temperatures}
\end{figure}

It is important to analyze the behavior of the phase diagram as
function of the gap parameter $M_0$. In Fig.\ref{Curie_Temperatures}
we show the phase diagram of a magnetically doped thick TI slab as a
function of $M_0$. For $M_0<0$ the system is a normal insulator and
there are no surface states. Also the spin orbit coupling is small
and the SDW phase does not exist. For $M_0>0$ the system is a TI and
the gap increases with $M_0$. TI with larger gaps have more metallic
surface states and the FM order at the surface is therefore more
robust. Also the effective spin orbit coupling is stronger and both
$T^*$ and $T_G$ increase with $M_0$. The results of
Fig.\ref{Curie_Temperatures} show that the range of temperatures
where the SDW phase exists increases with $M_0$.

Finally we make an estimation of the critical temperature. From the
band structure parameters of Bi$_2$Se$_3$, choosing the density of
the magnetic impurities to be 5$\times$10$^{20}$cm$^{-3}$, the total
angular momentum of a single magnetic ion to be $S=3/2$ and the
effective exchange coupling $J_{eff}$ =250$meV nm^3$\cite{Yu_2010}
we obtain $T_G ^{bulk} \approx$18$K$. This values can change by
factors of two by changing the  magnetic ions or the density of
impurities. It is well known that the mean-field approximations tend
to overestimate the transition temperature due to the neglect of the
fluctuations. In diluted magnetic semiconductors thermal
fluctuations reduce the value of the Curie temperature in near
30$\%$\cite{Brey_2003},and we expect a similar reduction in
topological insulators.

\acknowledgments We are very grateful to P.G.Silvestrov who call our
attention on the correct definition of the spin matrices. We also
acknowledge fruitful discussion with E.Chacón, H.A.Fertig and
A.H.MacDonald. Funding for this work was provided by MICINN-Spain
via grant FIS2009-08744.

\appendix
\section{Mean Field Expression for the Entropy.}

In this section we obtain an expression for the entropy of a system
of classical spins of magnetization $m$, that are coupled with the
topological insulator trough a general term $E[m]$.

The Free Energy of a system of classical spins of magnitude unity in
an external magnetic field $h$ is
\begin{equation}
\mathcal{F}= - \frac{1}{\beta}\ln \left ( 2 \frac{\sinh(\beta h
)}{\beta h } \right ) \,,
\end{equation}
from where the magnetization can be calculated as
\begin{equation}
m \equiv <m> = - \frac{\partial \mathcal{F} }{\partial h } =
\frac{1}{\tanh ( \beta h)} -\frac{1}{\beta h} \,.
\label{magnetization}
\end{equation}

The entropy of the spin system is then
\begin{equation}
-TS= \mathcal{F} - m h = \frac{1}{\beta} \left ( - \ln \left ( 2
\frac{\sinh ( \beta h )}{\beta h } \right )  - m \beta h  \right )
\,.
\end{equation}
The total energy of the system  is
\begin{equation}
\mathcal{F} ^{total} =E[m]-TS=E[m]-\frac 1 {\beta}\left [\ln \left (
2 \frac {\sinh (\beta h)} {\beta h} \right ) + m \beta h  \right ]
\end{equation}
where $E[m]$ is the change in the electronic energy of the system
because of the polarization of the magnetic impurities.

To obtain $h$, we minimize the total free energy with respect to
$h$, $\partial \mathcal{F} ^{\rm total}/\partial h =0$. In the limit
of small $h$
\begin{equation}
\ln ( 2 \frac{\sinh ( \beta h ) }{\beta h })  \simeq \ln ( 2) +
\frac{( \beta h ) ^2 }{6}- \frac {(\beta h ) ^4}{180} + ...
\end{equation}
In this limit, the minimization condition gives $\beta h = -3
m-\frac 3 5 m^3+...$, and the entropy gets the form
\begin{equation}
-TS = - k_B T  \ln (2)  + \frac{3}{2}k_{\rm B} T  m ^2 +
\frac{9}{20}k_{\rm B} T  m ^4.
\end{equation}

%
%
%


\begin{thebibliography}{40}
\expandafter\ifx\csname
natexlab\endcsname\relax\def\natexlab#1{#1}\fi
\expandafter\ifx\csname bibnamefont\endcsname\relax
  \def\bibnamefont#1{#1}\fi
\expandafter\ifx\csname bibfnamefont\endcsname\relax
  \def\bibfnamefont#1{#1}\fi
\expandafter\ifx\csname citenamefont\endcsname\relax
  \def\citenamefont#1{#1}\fi
\expandafter\ifx\csname url\endcsname\relax
  \def\url#1{\texttt{#1}}\fi
\expandafter\ifx\csname urlprefix\endcsname\relax\def\urlprefix{URL
}\fi \providecommand{\bibinfo}[2]{#2}
\providecommand{\eprint}[2][]{\url{#2}}

\bibitem[{\citenamefont{Hasan and Kane}(2010)}]{hasan_2010}
\bibinfo{author}{\bibfnamefont{M.~Z.} \bibnamefont{Hasan}} \bibnamefont{and}
  \bibinfo{author}{\bibfnamefont{C.~L.} \bibnamefont{Kane}},
  \bibinfo{journal}{Rev. Mod. Phys.} \textbf{\bibinfo{volume}{82}},
  \bibinfo{pages}{3045} (\bibinfo{year}{2010}).

\bibitem[{\citenamefont{Qi and Zhang}(2010)}]{Qi_2010}
\bibinfo{author}{\bibfnamefont{X.-L.} \bibnamefont{Qi}} \bibnamefont{and}
  \bibinfo{author}{\bibfnamefont{S.-C.} \bibnamefont{Zhang}},
  \bibinfo{journal}{Physics Today} \textbf{\bibinfo{volume}{63}},
  \bibinfo{pages}{33} (\bibinfo{year}{2010}).

\bibitem[{\citenamefont{Qi and Zhang}(2011)}]{Qi_2011}
\bibinfo{author}{\bibfnamefont{X.-L.} \bibnamefont{Qi}} \bibnamefont{and}
  \bibinfo{author}{\bibfnamefont{S.-C.} \bibnamefont{Zhang}},
  \bibinfo{journal}{Rev. Mod. Phys.} \textbf{\bibinfo{volume}{83}},
  \bibinfo{pages}{1057} (\bibinfo{year}{2011}).

\bibitem[{\citenamefont{{L. Xia {\it et al.}}}(2009)}]{Xia_2009}
\bibinfo{author}{\bibnamefont{{L. Xia {\it et al.}}}}, \bibinfo{journal}{Nature
  Phys.} \textbf{\bibinfo{volume}{5}}, \bibinfo{pages}{398}
  (\bibinfo{year}{2009}).

\bibitem[{\citenamefont{{H. Zhang, {\it et al.}}}(2009)}]{Zhang_2009}
\bibinfo{author}{\bibnamefont{{H. Zhang, {\it et al.}}}},
  \bibinfo{journal}{Nature Phys.} \textbf{\bibinfo{volume}{5}},
  \bibinfo{pages}{438} (\bibinfo{year}{2009}).

\bibitem[{\citenamefont{{D. Hsieh {\it et al.} }}(2009)}]{Hsieh_2009}
\bibinfo{author}{\bibnamefont{{D. Hsieh {\it et al.} }}},
  \bibinfo{journal}{Nature} \textbf{\bibinfo{volume}{460}},
  \bibinfo{pages}{1101} (\bibinfo{year}{2009}).

\bibitem[{\citenamefont{Hanaguri et~al.}(2010)\citenamefont{Hanaguri, Igarashi,
  Kawamura, Takagi, and Sasagawa}}]{Hanaguri_2010}
\bibinfo{author}{\bibfnamefont{T.}~\bibnamefont{Hanaguri}},
  \bibinfo{author}{\bibfnamefont{K.}~\bibnamefont{Igarashi}},
  \bibinfo{author}{\bibfnamefont{M.}~\bibnamefont{Kawamura}},
  \bibinfo{author}{\bibfnamefont{H.}~\bibnamefont{Takagi}}, \bibnamefont{and}
  \bibinfo{author}{\bibfnamefont{T.}~\bibnamefont{Sasagawa}},
  \bibinfo{journal}{Phys. Rev. B} \textbf{\bibinfo{volume}{82}},
  \bibinfo{pages}{081305} (\bibinfo{year}{2010}).

\bibitem[{\citenamefont{Qi et~al.}(2008)\citenamefont{Qi, Hughes, and
  Zhang}}]{Qi_2008}
\bibinfo{author}{\bibfnamefont{X.-L.} \bibnamefont{Qi}},
  \bibinfo{author}{\bibfnamefont{T.~L.} \bibnamefont{Hughes}},
  \bibnamefont{and} \bibinfo{author}{\bibfnamefont{S.-C.} \bibnamefont{Zhang}},
  \bibinfo{journal}{Phys. Rev. B} \textbf{\bibinfo{volume}{78}},
  \bibinfo{pages}{195424} (\bibinfo{year}{2008}).

\bibitem[{\citenamefont{Essin et~al.}(2009)\citenamefont{Essin, Moore, and
  Vanderbilt}}]{Essin_2009}
\bibinfo{author}{\bibfnamefont{A.~M.} \bibnamefont{Essin}},
  \bibinfo{author}{\bibfnamefont{J.~E.} \bibnamefont{Moore}}, \bibnamefont{and}
  \bibinfo{author}{\bibfnamefont{D.}~\bibnamefont{Vanderbilt}},
  \bibinfo{journal}{Phys. Rev. Lett.} \textbf{\bibinfo{volume}{102}},
  \bibinfo{pages}{146805} (\bibinfo{year}{2009}).

\bibitem[{\citenamefont{Liu et~al.}(2009)\citenamefont{Liu, Liu, Xu, Qi, and
  Zhang}}]{Liu_2009}
\bibinfo{author}{\bibfnamefont{Q.}~\bibnamefont{Liu}},
  \bibinfo{author}{\bibfnamefont{C.-X.} \bibnamefont{Liu}},
  \bibinfo{author}{\bibfnamefont{C.}~\bibnamefont{Xu}},
  \bibinfo{author}{\bibfnamefont{X.-L.} \bibnamefont{Qi}}, \bibnamefont{and}
  \bibinfo{author}{\bibfnamefont{S.-C.} \bibnamefont{Zhang}},
  \bibinfo{journal}{Phys. Rev. Lett.} \textbf{\bibinfo{volume}{102}},
  \bibinfo{pages}{156603} (\bibinfo{year}{2009}).

\bibitem[{\citenamefont{Biswas and Balatsky}(2010)}]{Biswas_2010}
\bibinfo{author}{\bibfnamefont{R.~R.} \bibnamefont{Biswas}} \bibnamefont{and}
  \bibinfo{author}{\bibfnamefont{A.~V.} \bibnamefont{Balatsky}},
  \bibinfo{journal}{Phys. Rev. B} \textbf{\bibinfo{volume}{81}},
  \bibinfo{pages}{233405} (\bibinfo{year}{2010}).

\bibitem[{\citenamefont{Yu et~al.}(2010)\citenamefont{Yu, Zhang, Zhang, Zhang,
  Dai, and Fang}}]{Yu_2010}
\bibinfo{author}{\bibfnamefont{R.}~\bibnamefont{Yu}},
  \bibinfo{author}{\bibfnamefont{W.}~\bibnamefont{Zhang}},
  \bibinfo{author}{\bibfnamefont{H.-J.} \bibnamefont{Zhang}},
  \bibinfo{author}{\bibfnamefont{S.-C.} \bibnamefont{Zhang}},
  \bibinfo{author}{\bibfnamefont{X.}~\bibnamefont{Dai}}, \bibnamefont{and}
  \bibinfo{author}{\bibfnamefont{Z.}~\bibnamefont{Fang}},
  \bibinfo{journal}{Science} \textbf{\bibinfo{volume}{329}},
  \bibinfo{pages}{61} (\bibinfo{year}{2010}).

\bibitem[{\citenamefont{Jiang and Wu}(2011)}]{Jiang_2011}
\bibinfo{author}{\bibfnamefont{J.-H.} \bibnamefont{Jiang}} \bibnamefont{and}
  \bibinfo{author}{\bibfnamefont{S.}~\bibnamefont{Wu}}, \bibinfo{journal}{Phys.
  Rev. B} \textbf{\bibinfo{volume}{83}}, \bibinfo{pages}{205124}
  (\bibinfo{year}{2011}).

\bibitem[{\citenamefont{Yokoyama}(2011)}]{Yokoyama_2011}
\bibinfo{author}{\bibfnamefont{T.}~\bibnamefont{Yokoyama}},
  \bibinfo{journal}{Phys. Rev. B} \textbf{\bibinfo{volume}{84}},
  \bibinfo{pages}{113407} (\bibinfo{year}{2011}).

\bibitem[{\citenamefont{Zhu et~al.}(2011)\citenamefont{Zhu, Yao, Zhang, and
  Chang}}]{Zhu_2011}
\bibinfo{author}{\bibfnamefont{J.-J.} \bibnamefont{Zhu}},
  \bibinfo{author}{\bibfnamefont{D.-X.} \bibnamefont{Yao}},
  \bibinfo{author}{\bibfnamefont{S.-C.} \bibnamefont{Zhang}}, \bibnamefont{and}
  \bibinfo{author}{\bibfnamefont{K.}~\bibnamefont{Chang}},
  \bibinfo{journal}{Phys. Rev. Lett.} \textbf{\bibinfo{volume}{106}},
  \bibinfo{pages}{097201} (\bibinfo{year}{2011}).

\bibitem[{\citenamefont{Abanin and Pesin}(2011)}]{Abanin_2011}
\bibinfo{author}{\bibfnamefont{D.~A.} \bibnamefont{Abanin}} \bibnamefont{and}
  \bibinfo{author}{\bibfnamefont{D.~A.} \bibnamefont{Pesin}},
  \bibinfo{journal}{Phys. Rev. Lett.} \textbf{\bibinfo{volume}{106}},
  \bibinfo{pages}{136802} (\bibinfo{year}{2011}).

\bibitem[{\citenamefont{Men'shov et~al.}(2011)\citenamefont{Men'shov, Tugushev,
  and Chulkov}}]{Menshov_2011}
\bibinfo{author}{\bibfnamefont{V.~N.} \bibnamefont{Men'shov}},
  \bibinfo{author}{\bibfnamefont{V.}~\bibnamefont{Tugushev}}, \bibnamefont{and}
  \bibinfo{author}{\bibfnamefont{E.}~\bibnamefont{Chulkov}},
  \bibinfo{journal}{JETP Letters} \textbf{\bibinfo{volume}{94}},
  \bibinfo{pages}{629} (\bibinfo{year}{2011}).

\bibitem[{\citenamefont{Nunez and Fern\'andez-Rossier}(2012)}]{nunez_2012}
\bibinfo{author}{\bibfnamefont{A.}~\bibnamefont{Nunez}} \bibnamefont{and}
  \bibinfo{author}{\bibfnamefont{J.}~\bibnamefont{Fern\'andez-Rossier}},
  \bibinfo{journal}{Solid State Communications} \textbf{\bibinfo{volume}{152}},
  \bibinfo{pages}{403 } (\bibinfo{year}{2012}).

\bibitem[{\citenamefont{Habe and Asano}(2012)}]{Tetsuro_2012}
\bibinfo{author}{\bibfnamefont{T.}~\bibnamefont{Habe}} \bibnamefont{and}
  \bibinfo{author}{\bibfnamefont{Y.}~\bibnamefont{Asano}},
  \bibinfo{journal}{Phys. Rev. B} \textbf{\bibinfo{volume}{85}},
  \bibinfo{pages}{195325} (\bibinfo{year}{2012}).

\bibitem[{\citenamefont{{ Y.L. Chen {\it et al.}}}(2010)}]{Chen_2010}
\bibinfo{author}{\bibnamefont{{ Y.L. Chen {\it et al.}}}},
  \bibinfo{journal}{Science} \textbf{\bibinfo{volume}{329}},
  \bibinfo{pages}{659} (\bibinfo{year}{2010}).

\bibitem[{\citenamefont{{L.A. Wray {\it et al.} }}(2010)}]{Wray_2010}
\bibinfo{author}{\bibnamefont{{L.A. Wray {\it et al.} }}},
  \bibinfo{journal}{Nature Phys.} \textbf{\bibinfo{volume}{7}},
  \bibinfo{pages}{32} (\bibinfo{year}{2010}).

\bibitem[{\citenamefont{{Chang} et~al.}(2011)\citenamefont{{Chang}, {Zhang},
  {Liu}, {Zhang}, {Feng}, {Li}, {Wang}, {Chen}, {Dai}, {Fang}
  et~al.}}]{Chang_2011}
\bibinfo{author}{\bibfnamefont{C.-Z.} \bibnamefont{{Chang}}},
  \bibinfo{author}{\bibfnamefont{J.-S.} \bibnamefont{{Zhang}}},
  \bibinfo{author}{\bibfnamefont{M.-H.} \bibnamefont{{Liu}}},
  \bibinfo{author}{\bibfnamefont{Z.-C.} \bibnamefont{{Zhang}}},
  \bibinfo{author}{\bibfnamefont{X.}~\bibnamefont{{Feng}}},
  \bibinfo{author}{\bibfnamefont{K.}~\bibnamefont{{Li}}},
  \bibinfo{author}{\bibfnamefont{L.-L.} \bibnamefont{{Wang}}},
  \bibinfo{author}{\bibfnamefont{X.}~\bibnamefont{{Chen}}},
  \bibinfo{author}{\bibfnamefont{X.}~\bibnamefont{{Dai}}},
  \bibinfo{author}{\bibfnamefont{Z.}~\bibnamefont{{Fang}}},
  \bibnamefont{et~al.}, \bibinfo{journal}{ArXiv e-prints}
  (\bibinfo{year}{2011}), \eprint{1108.4754}.

\bibitem[{\citenamefont{{Honolka} et~al.}(2011)\citenamefont{{Honolka},
  {Khajetoorians}, {Sessi}, {Wehling}, {Stepanow}, {Mi}, {Iversen}, {Schlenk},
  {Wiebe}, {Brookes} et~al.}}]{Honolka_2011}
\bibinfo{author}{\bibfnamefont{J.}~\bibnamefont{{Honolka}}},
  \bibinfo{author}{\bibfnamefont{A.~A.} \bibnamefont{{Khajetoorians}}},
  \bibinfo{author}{\bibfnamefont{V.}~\bibnamefont{{Sessi}}},
  \bibinfo{author}{\bibfnamefont{T.~O.} \bibnamefont{{Wehling}}},
  \bibinfo{author}{\bibfnamefont{S.}~\bibnamefont{{Stepanow}}},
  \bibinfo{author}{\bibfnamefont{J.-L.} \bibnamefont{{Mi}}},
  \bibinfo{author}{\bibfnamefont{B.~B.} \bibnamefont{{Iversen}}},
  \bibinfo{author}{\bibfnamefont{T.}~\bibnamefont{{Schlenk}}},
  \bibinfo{author}{\bibfnamefont{J.}~\bibnamefont{{Wiebe}}},
  \bibinfo{author}{\bibfnamefont{N.}~\bibnamefont{{Brookes}}},
  \bibnamefont{et~al.}, \bibinfo{journal}{ArXiv e-prints}
  (\bibinfo{year}{2011}), \eprint{1112.4621}.

\bibitem[{\citenamefont{{Scholz} et~al.}(2011)\citenamefont{{Scholz},
  {S{\'a}nchez-Barriga}, {Marchenko}, {Varykhalov}, {Volykhov}, {Yashina}, and
  {Rader}}}]{Scholz_2011}
\bibinfo{author}{\bibfnamefont{M.~R.} \bibnamefont{{Scholz}}},
  \bibinfo{author}{\bibfnamefont{J.}~\bibnamefont{{S{\'a}nchez-Barriga}}},
  \bibinfo{author}{\bibfnamefont{D.}~\bibnamefont{{Marchenko}}},
  \bibinfo{author}{\bibfnamefont{A.}~\bibnamefont{{Varykhalov}}},
  \bibinfo{author}{\bibfnamefont{A.}~\bibnamefont{{Volykhov}}},
  \bibinfo{author}{\bibfnamefont{L.~V.} \bibnamefont{{Yashina}}},
  \bibnamefont{and} \bibinfo{author}{\bibfnamefont{O.}~\bibnamefont{{Rader}}},
  \bibinfo{journal}{ArXiv e-prints}  (\bibinfo{year}{2011}),
  \eprint{1108.1037}.

\bibitem[{\citenamefont{Valla et~al.}(2012)\citenamefont{Valla, Pan, Gardner,
  Lee, and Chu}}]{Valla_2012}
\bibinfo{author}{\bibfnamefont{T.}~\bibnamefont{Valla}},
  \bibinfo{author}{\bibfnamefont{Z.-H.} \bibnamefont{Pan}},
  \bibinfo{author}{\bibfnamefont{D.}~\bibnamefont{Gardner}},
  \bibinfo{author}{\bibfnamefont{Y.~S.} \bibnamefont{Lee}}, \bibnamefont{and}
  \bibinfo{author}{\bibfnamefont{S.}~\bibnamefont{Chu}},
  \bibinfo{journal}{Phys. Rev. Lett.} \textbf{\bibinfo{volume}{108}},
  \bibinfo{pages}{117601} (\bibinfo{year}{2012}).

\bibitem[{\citenamefont{{Xu} et~al.}(2012)\citenamefont{{Xu}, {Wray},
  {Alidoust}, {Xia}, {Neupane}, {Liu}, {Ji}, {Jia}, {Cava}, and
  {Hasan}}}]{Xu_2012}
\bibinfo{author}{\bibfnamefont{S.-Y.} \bibnamefont{{Xu}}},
  \bibinfo{author}{\bibfnamefont{L.~A.} \bibnamefont{{Wray}}},
  \bibinfo{author}{\bibfnamefont{N.}~\bibnamefont{{Alidoust}}},
  \bibinfo{author}{\bibfnamefont{Y.}~\bibnamefont{{Xia}}},
  \bibinfo{author}{\bibfnamefont{M.}~\bibnamefont{{Neupane}}},
  \bibinfo{author}{\bibfnamefont{C.}~\bibnamefont{{Liu}}},
  \bibinfo{author}{\bibfnamefont{H.-W.} \bibnamefont{{Ji}}},
  \bibinfo{author}{\bibfnamefont{S.}~\bibnamefont{{Jia}}},
  \bibinfo{author}{\bibfnamefont{R.~J.} \bibnamefont{{Cava}}},
  \bibnamefont{and} \bibinfo{author}{\bibfnamefont{M.~Z.}
  \bibnamefont{{Hasan}}}, \bibinfo{journal}{ArXiv e-prints}
  (\bibinfo{year}{2012}), \eprint{1206.0278}.

\bibitem[{\citenamefont{Schmidt et~al.}(2011)\citenamefont{Schmidt, Miwa, and
  Fazzio}}]{Schimdt_2011}
\bibinfo{author}{\bibfnamefont{T.~M.} \bibnamefont{Schmidt}},
  \bibinfo{author}{\bibfnamefont{R.~H.} \bibnamefont{Miwa}}, \bibnamefont{and}
  \bibinfo{author}{\bibfnamefont{A.}~\bibnamefont{Fazzio}},
  \bibinfo{journal}{Phys. Rev. B} \textbf{\bibinfo{volume}{84}},
  \bibinfo{pages}{245418} (\bibinfo{year}{2011}).

\bibitem[{\citenamefont{Liu et~al.}(2010)\citenamefont{Liu, Qi, Zhang, Dai,
  Fang, and Zhang}}]{Liu_2010}
\bibinfo{author}{\bibfnamefont{C.-X.} \bibnamefont{Liu}},
  \bibinfo{author}{\bibfnamefont{X.-L.} \bibnamefont{Qi}},
  \bibinfo{author}{\bibfnamefont{H.}~\bibnamefont{Zhang}},
  \bibinfo{author}{\bibfnamefont{X.}~\bibnamefont{Dai}},
  \bibinfo{author}{\bibfnamefont{Z.}~\bibnamefont{Fang}}, \bibnamefont{and}
  \bibinfo{author}{\bibfnamefont{S.-C.} \bibnamefont{Zhang}},
  \bibinfo{journal}{Phys. Rev. B} \textbf{\bibinfo{volume}{82}},
  \bibinfo{pages}{045122} (\bibinfo{year}{2010}).

\bibitem[{\citenamefont{P.G.Silvestrov
  et~al.}(2011)\citenamefont{P.G.Silvestrov, P.W.Brouwer, and
  E.G.Mishchenlo}}]{Silvestrov_2011}
\bibinfo{author}{\bibnamefont{P.G.Silvestrov}},
  \bibinfo{author}{\bibnamefont{P.W.Brouwer}}, \bibnamefont{and}
  \bibinfo{author}{\bibnamefont{E.G.Mishchenlo}} (\bibinfo{year}{2011}),
  \urlprefix\url{arXiv.org:1111.3650}.

\bibitem[{\citenamefont{Vleck}(1932)}]{Vleck_1932}
\bibinfo{author}{\bibfnamefont{J.}~\bibnamefont{Vleck}},
  \emph{\bibinfo{title}{The Theory of Electronic and Magnetic
  Susceptibilities.}} (\bibinfo{publisher}{Oxford University Press, London},
  \bibinfo{year}{1932}).

\bibitem[{\citenamefont{Hor et~al.}(2009)\citenamefont{Hor, Richardella,
  Roushan, Xia, Checkelsky, Yazdani, Hasan, Ong, and Cava}}]{Hor_2009}
\bibinfo{author}{\bibfnamefont{Y.~S.} \bibnamefont{Hor}},
  \bibinfo{author}{\bibfnamefont{A.}~\bibnamefont{Richardella}},
  \bibinfo{author}{\bibfnamefont{P.}~\bibnamefont{Roushan}},
  \bibinfo{author}{\bibfnamefont{Y.}~\bibnamefont{Xia}},
  \bibinfo{author}{\bibfnamefont{J.~G.} \bibnamefont{Checkelsky}},
  \bibinfo{author}{\bibfnamefont{A.}~\bibnamefont{Yazdani}},
  \bibinfo{author}{\bibfnamefont{M.~Z.} \bibnamefont{Hasan}},
  \bibinfo{author}{\bibfnamefont{N.~P.} \bibnamefont{Ong}}, \bibnamefont{and}
  \bibinfo{author}{\bibfnamefont{R.~J.} \bibnamefont{Cava}},
  \bibinfo{journal}{Phys. Rev. B} \textbf{\bibinfo{volume}{79}},
  \bibinfo{pages}{195208} (\bibinfo{year}{2009}).

\bibitem[{\citenamefont{Dietl et~al.}(2000)\citenamefont{Dietl, Ohno,
  Matsukura, Cibert, and Ferrand}}]{Dietl_2000}
\bibinfo{author}{\bibfnamefont{T.}~\bibnamefont{Dietl}},
  \bibinfo{author}{\bibfnamefont{H.}~\bibnamefont{Ohno}},
  \bibinfo{author}{\bibfnamefont{F.}~\bibnamefont{Matsukura}},
  \bibinfo{author}{\bibfnamefont{J.}~\bibnamefont{Cibert}}, \bibnamefont{and}
  \bibinfo{author}{\bibfnamefont{D.}~\bibnamefont{Ferrand}},
  \bibinfo{journal}{Science} \textbf{\bibinfo{volume}{287}},
  \bibinfo{pages}{1019} (\bibinfo{year}{2000}).

\bibitem[{\citenamefont{Brey and G\'omez-Santos}(2003)}]{Brey_2003}
\bibinfo{author}{\bibfnamefont{L.}~\bibnamefont{Brey}} \bibnamefont{and}
  \bibinfo{author}{\bibfnamefont{G.}~\bibnamefont{G\'omez-Santos}},
  \bibinfo{journal}{Phys. Rev. B} \textbf{\bibinfo{volume}{68}},
  \bibinfo{pages}{115206} (\bibinfo{year}{2003}).

\bibitem[{\citenamefont{Calder\'on et~al.}(2002)\citenamefont{Calder\'on,
  G\'omez-Santos, and Brey}}]{Calderon_2002}
\bibinfo{author}{\bibfnamefont{M.~J.} \bibnamefont{Calder\'on}},
  \bibinfo{author}{\bibfnamefont{G.}~\bibnamefont{G\'omez-Santos}},
  \bibnamefont{and} \bibinfo{author}{\bibfnamefont{L.}~\bibnamefont{Brey}},
  \bibinfo{journal}{Phys. Rev. B} \textbf{\bibinfo{volume}{66}},
  \bibinfo{pages}{075218} (\bibinfo{year}{2002}).

\bibitem[{\citenamefont{Abolfath et~al.}(2001)\citenamefont{Abolfath,
  Jungwirth, Brum, and MacDonald}}]{Abolfath_2001}
\bibinfo{author}{\bibfnamefont{M.}~\bibnamefont{Abolfath}},
  \bibinfo{author}{\bibfnamefont{T.}~\bibnamefont{Jungwirth}},
  \bibinfo{author}{\bibfnamefont{J.}~\bibnamefont{Brum}}, \bibnamefont{and}
  \bibinfo{author}{\bibfnamefont{A.~H.} \bibnamefont{MacDonald}},
  \bibinfo{journal}{Phys. Rev. B} \textbf{\bibinfo{volume}{63}},
  \bibinfo{pages}{054418} (\bibinfo{year}{2001}).

\bibitem[{\citenamefont{Fern\'andez-Rossier and Sham}(2001)}]{FR_2001}
\bibinfo{author}{\bibfnamefont{J.}~\bibnamefont{Fern\'andez-Rossier}}
  \bibnamefont{and} \bibinfo{author}{\bibfnamefont{L.~J.} \bibnamefont{Sham}},
  \bibinfo{journal}{Phys. Rev. B} \textbf{\bibinfo{volume}{64}},
  \bibinfo{pages}{235323} (\bibinfo{year}{2001}).

\bibitem[{\citenamefont{P.M.Chaikin and T.C.Lubensky}(1995)}]{chaikin_book}
\bibinfo{author}{\bibnamefont{P.M.Chaikin}} \bibnamefont{and}
  \bibinfo{author}{\bibnamefont{T.C.Lubensky}},
  \emph{\bibinfo{title}{Principles of Condensed Matter Physics}}
  (\bibinfo{publisher}{Cambridge University Press},
  \bibinfo{address}{Cambridge, UK}, \bibinfo{year}{1995}).

\bibitem[{\citenamefont{Brey et~al.}(2007)\citenamefont{Brey, Fertig, and
  Das~Sarma}}]{Brey_2007}
\bibinfo{author}{\bibfnamefont{L.}~\bibnamefont{Brey}},
  \bibinfo{author}{\bibfnamefont{H.~A.} \bibnamefont{Fertig}},
  \bibnamefont{and}
  \bibinfo{author}{\bibfnamefont{S.}~\bibnamefont{Das~Sarma}},
  \bibinfo{journal}{Phys. Rev. Lett.} \textbf{\bibinfo{volume}{99}},
  \bibinfo{pages}{116802} (\bibinfo{year}{2007}).

\bibitem[{\citenamefont{Rosenberg and Franz}(2012)}]{Rosenberg_2012}
\bibinfo{author}{\bibfnamefont{G.}~\bibnamefont{Rosenberg}} \bibnamefont{and}
  \bibinfo{author}{\bibfnamefont{M.}~\bibnamefont{Franz}},
  \bibinfo{journal}{Phys. Rev. B} \textbf{\bibinfo{volume}{85}},
  \bibinfo{pages}{195119} (\bibinfo{year}{2012}).

\bibitem[{\citenamefont{{Salman} et~al.}(2012)\citenamefont{{Salman},
  {Pomjakushina}, {Pomjakushin}, {Kanigel}, {Chashka}, {Conder}, {Morenzoni},
  {Prokscha}, {Sedlak}, and {Suter}}}]{Salman_2012}
\bibinfo{author}{\bibfnamefont{Z.}~\bibnamefont{{Salman}}},
  \bibinfo{author}{\bibfnamefont{E.}~\bibnamefont{{Pomjakushina}}},
  \bibinfo{author}{\bibfnamefont{V.}~\bibnamefont{{Pomjakushin}}},
  \bibinfo{author}{\bibfnamefont{A.}~\bibnamefont{{Kanigel}}},
  \bibinfo{author}{\bibfnamefont{K.}~\bibnamefont{{Chashka}}},
  \bibinfo{author}{\bibfnamefont{K.}~\bibnamefont{{Conder}}},
  \bibinfo{author}{\bibfnamefont{E.}~\bibnamefont{{Morenzoni}}},
  \bibinfo{author}{\bibfnamefont{T.}~\bibnamefont{{Prokscha}}},
  \bibinfo{author}{\bibfnamefont{K.}~\bibnamefont{{Sedlak}}}, \bibnamefont{and}
  \bibinfo{author}{\bibfnamefont{A.}~\bibnamefont{{Suter}}},
  \bibinfo{journal}{ArXiv e-prints}  (\bibinfo{year}{2012}),
  \eprint{1203.4850}.

\end{thebibliography}

\end{document}